\documentstyle[12pt]{article}
\begin{document}
\renewcommand{\theequation}{\arabic{section}.\arabic{equation}}
\pagestyle{plain} \input psfig 
\begin{titlepage} \vspace*{-2cm} Preprint \hfill
\hbox{\bf SB/F/96-235} \hrule 
\vskip 3cm 
\centerline{ \Large{\bf Magnetic monopoles over topologically}} 
\centerline{ \Large{\bf non trivial Riemann Surfaces}}
\vskip 1cm

\centerline{ I. Martin and A. Restuccia} \vskip 4mm

\centerline{\it Universidad Sim\'on Bol\'{\i}var,Departamento de F\'{\i}sica}
\centerline{\it Apartado Postal 89000, Caracas 1080-A, Venezuela.}

\vskip 3.5cm

{\bf Abstract} An explicit canonical construction of monopole connections on
non
trivial $U(1)$ bundles over Riemann surfaces of any genus is given. The class
of
monopole solutions depend on the conformal class of the given Riemann surface
and a set of integer weights. The reduction of Seiberg-Witten 4-monopole
equations to Riemann surfaces is performed. It is shown then that the monopole
connections constructed are solutions to these equations. \vskip 1.5cm

\vspace*{4cm} \hrule \bigskip \centerline{\footnotesize {\it{ e-mail:
isbeliam@usb.ve , arestu@usb.ve} }} \vfill \end{titlepage}

\section{ {Introduction}}

\hspace{.9cm}	The duality symmetry, associated to a $U(1)$ gauge symmetry,
is one of the symmetries that lately has risen high expectations in QFT mainly
because it offers a way of analysing field theories describing strong-coupling
interaction and weak-coupling interactions on an equal footing. In fact,
recently Seiberg and Witten [1] obtained the full effective action for the
fields at any coupling, from knowledge of its weak and strong coupling limit,
considering the breaking of the $SU(2)$ gauge down to $U(1)$ in the $N=2$
supersymmetric $SU(2)$ Yang Mills theory.

It has long been known that the monopole solutions i.e solutions to some vacum
equations in QFT with a "magnetic charge" different from zero always appear
whenever there is a breaking of a compact gauge symmetry down to a $U(1)$ gauge
[2]. These solutions resemble Dirac monopoles at infinity. The Seiberg-Witten
procedure mentioned before gave rise to some new equations where monopoles play
a significant role once again. These Seiberg-Witten equations have also
interested the mathematics community mainly because, as a Topological QFT, they
give some new topological invariants that may provide a new approach to the
classification of 4-manifolds given by the Donaldson invariants. Most recently,
it has been proven [3] an equivalence between Seiberg-Witten invariants of a
symplectic 4-manifold and a set of Gromov invariants defined using
pseudo-holomorphic submanifolds of dimension 2.

In this article, we give a canonical construction of monopole connections of a
non trivial $U(1)$ bundle over Riemann surfaces of any genus.The solutions
constructed here should appear as soliton solutions to the low energy heterotic
string field equations over non trivial backgrounds [4]. They may provide some
insight on the non-perturbative semiclassical structure of strings and membrane
theories. Also, in this article, we project the Seiberg-Witten equations
locally
to the complex plane, impose conditions on the components of the fields normal
to the plane and then extend them globally to build up the equations over
compact Riemann surfaces. We end up showing that the canonical monopole
connections obtained earlier are solutions of the projected Seiberg-Witten
equations over compact Riemann surfaces. We expect these solutions, following
the ideas of Taubes [3], to be 'grafted' into solutions of the Seiberg-Witten
equations over a not necessarily symplectic 4-manifold. They also may help in
the search of Gromov type invariants using pseudo-holomorphic submanifolds with
degenerate symplectic forms.

In section 2, we give a construction of the known Dirac monopole connection
mainly to compare it with the new solutions over Riemann surfaces of any genus
given in section 3 and show that our solutions include the Dirac monopole for
genus zero. In section 4, we prove that the new monopole connections are
solutions to the reduced Seiberg-Witten equations over compact Riemann
surfaces.
\section{ Dirac monopoles over $S_2$} \hspace{.9cm}We first describe the
Hopf fiber bundle over $S_2$ and  the connection \linebreak 1-form and
curvatures 2-form of the Dirac monopole. In the next section we generalize the
solutions to any compact Riemann surface.

The 3-dimensional sphere $S_3$ may be defined  by $z_0, z_1 \in C
\hspace{-7.5pt} l$, the complex numbers, satisfying

\begin{equation} z_0\bar{z}_0+z_1\bar{z}_1=1 \end{equation}

The group $U(1)$ acts on $S_3$ by

\begin{equation} (z_0,z_1) \longrightarrow (z_0u,z_1u) \end{equation} where
$u\bar{u}=1,\bar{u}$ defines the complex conjugate to $u\in C \hspace{-7.5pt}
l$

The projection $S_3{\longrightarrow}S_2$ is defined by the composition of

\begin{equation} (z_0,z_1) \longrightarrow \left\{\begin{array}{cc} z_1/z_0 &
z_0\neq0\\ z_0/z_1 & z_1\neq0 \end{array} \right. \end{equation} with the
stereographic projection \begin{equation} C \hspace{-7.5pt}l {\longrightarrow}
S_2 \end{equation} such that for

\begin{equation} z \in C \hspace{-7.5pt} l \: , \hspace{1cm} z=\rho e^{i\phi}
\end{equation} the stereographic projection associates $(\theta,\phi)$ over
$S_2$ with

\begin{equation} \rho =\frac{sin(\theta)}{1-cos(\theta)} \end{equation}

Over the Hopf  fiber bundle, there is a natural connection which may be
obtained
from the line element of $S_3$

\begin{equation} ds^2= 4(d\bar{z}_0 dz_0 + d\bar{z}_1 dz_1) \end{equation} this
can be decomposed in an unique way into the line element of $S_2$ and the
tensorial square of the 1-form

\begin{equation} \omega=d\chi + \cos(\theta) d\phi \end{equation} That is

\begin{equation} ds^2= (d{\theta}^2+\sin^2(\theta)d{\phi}^2) + {\omega}^2
\end{equation} where the Euler angles have been used

\begin{equation} \begin{array}{cc}
&z_0=[exp{\frac{1}{2}i(\chi+\phi)}]\cos({\theta}/2)\\ \\
&z_1=[exp{\frac{1}{2}i(\chi-\phi)}]\sin({\theta}/2) \end{array} \end{equation}
$\frac{1}{2}\omega$ defines a connection over  the fiber bundle $S_3$. To
obtain
the $U(1)$ connection 1-form over $S_2$, one may consider  the local section

\begin{equation} \hat{z}_0=e^{i\phi} \cos({\theta}/2), \hspace{1cm}
\hat{z}_1=\sin({\theta}/2) \end{equation} over $S_2$ with the point $\theta =0$
removed, which we denote $U_+$. The $U(1)$ connection 1-form over $U_+$ is
then\\ \begin{equation} A_+=\frac{1}{2} (1+\cos({\theta}))d\phi \end{equation}
which is regular on $U_+$.

If instead one considers the local section \begin{equation}
\tilde{z}_0=\cos({\theta}/2), \hspace{1cm}
\tilde{z}_1=e^{-i\phi}\sin({\theta}/2) \end{equation} over $S_2$ with the point
$\theta=\pi$ removed, denoted $U_-$, we obtain \begin{equation} A_-=\frac{1}{2}
(-1+\cos({\theta}))d\phi \end{equation} regular on $U_-$

In the overlapling region $U_+ \cap U_-$, we have

\begin{equation} (\tilde{z}_0,\tilde{z}_1)= ( e^{-i\phi} \hat{z}_0,
e^{-i\phi}\hat{z}_1) \end{equation} according to the action of $U(1)$ on the
fibers over $S_2$ and

\begin{equation} A_+=A_- + d\phi \end{equation}

The  curvature 2-form $\Omega$ arising from (12) and (14) is

\begin{equation} \Omega =\frac{1}{2} \sin(\theta) d\phi \wedge d\theta
\end{equation}

Notice  that  by applying an exterior derivative to (2.12) one obtains in
addition to (2.17) a $\delta$-function at $\theta=\pi$ from $dd\phi$ not being
defined on this point, however, it is annihilated because its coefficient
becomes zero at $\theta=\pi$. The same occurs with (2.14).

This construction of the Dirac monopole with charge $g=1/2$ may be generalized
[5] to obtain other nontrivial Hopf fibring over $S_2$ and curvatures
associated
to monopoles with charge g=n/2. In [5] the $U(1)$ Hopf fibring

\begin{equation} S_{2n+1} \longrightarrow C \hspace{-7.5pt}l \hspace{1.5pt}P_n
\end{equation} with base manifold $ C \hspace{-7.5pt}l \hspace{1.5pt}P_n$ and
fiber bundle space $S_{2n+1}$ was considered.

\begin{eqnarray*} z_0, z_1,....,z_n \in C \hspace{-7.5pt}l \nonumber
\end{eqnarray*} satisfying \begin{equation} z_0\bar{z}_0+z_1 \bar{z}_1+ \cdots
+
z_n  \bar{z}_n = 1 \end{equation} define the 2n+1-dimensional sphere
$S_{2n+1}$.

Let us introduce

\begin{equation} \xi_a= {\displaystyle \frac{z_a}{z_0}} \:\:\:\: ,z_0=\rho
e^{i\chi}, \:\:\:\:a=1 \cdots n \end{equation} where $z_0 \neq 0$, we then have
\begin{equation} {\rho}^2[1+\xi_a \bar {\xi}_a]=1 . \end{equation}

The line element of $S_{2n+1}$

\begin{equation} ds^2= dz_0d\bar{z}_0 +dz_ad\bar{z}_a \end{equation} may then
be
rewritten in an unique  way as

\begin{equation} ds^2= ( \rho^2 \delta_{ab} - \rho^4 \bar {\xi}_a \xi_b )d\xi_a
d\bar{\xi}_b + \omega^2 \end{equation} where \begin{equation} \omega= d\chi +
\frac{i}{2}\rho^2(\xi_a d\bar{\xi}_a - \bar {\xi}_a d\xi_b) \end{equation} and
the first term on the right hand  side of (2.24) is a positive definite line
element on $C \hspace{-7.5pt}l \hspace{2pt}P_n$, defining a metric which we
denote $h_{a\bar{b}}$. The curvature 2-form over $C\hspace{-7.5pt}l
\hspace{2pt}
P_n$ is then given by

\begin{equation} \Omega= ({\rho}^2 \delta_{ab} -{\rho}^4 \bar{\xi}_a \xi_b )i
d\xi_a \wedge d\bar{\xi}_b. \end{equation}

It is a solution of Maxwell  equations because

\begin{eqnarray} & \underbrace{\Omega \wedge \cdots \wedge \Omega}_{n-
factors}={h_{a_1 \bar{b}_1}....h_{a_n \bar{b}_n} d\xi_{a_1} \wedge
d\bar\xi_{b_1} \wedge \cdots} \nonumber \\ &={(det h)^{\frac{1}{2}} d\xi_1
\wedge \cdots \wedge d\xi_n \wedge d\bar{\xi}_1 \wedge \cdots d\bar{\xi}_n}\\
&\mbox{=volume element ,\hspace{3.7cm}}\nonumber \end{eqnarray} hence

\begin{equation} ^\ast \Omega \approx \underbrace{\Omega \wedge \cdots \wedge
\Omega}_{(n-1)- factors} \end{equation} and then

\begin{equation} d {^\ast \Omega}= 0 \end{equation} We may now consider as in
[5] embedding $C\hspace{-7.5pt}l \hspace{2pt} P_1$ into $C\hspace{-7.5pt}l
\hspace{2pt} P_n$

\begin{equation} \begin{array}{ccrc} &\xi_1=\left({\begin{array}{cr} n \\ 1
\end{array} } \right )^{\frac 12}\xi \\

&\xi_2=\left( {\begin{array}{cr} n \\ 2 \end{array} } \right )^{\frac 12}\xi^2
\\ & \vdots \\

&\xi_n= \left( {\begin{array}{cr} n \\ n \end{array} } \right )^{\frac 12}\xi
^n
\end{array} \end{equation}

After substitution of (2.29)  into (2.25) and several calculations, we obtain

\begin{equation} \Omega _n =n(1+ \xi \bar{\xi})^{-2}i d\xi \wedge d\bar{\xi}
\end{equation}

For $n=1$ we get

\begin{equation} \Omega_1= \frac{1}{(1+ \xi \bar{\xi})^2} i d\xi \wedge
d\bar{\xi} \end{equation} and by changing variables

\begin{equation} \xi=\rho e^{i\phi} \: , \: \rho=\frac{\sin \theta}{1- \cos
\theta} \end{equation} as in (2.6), \begin{equation} \begin{array}{rc}
\Omega_1&={\displaystyle \frac{2}{(1+\rho^2)^2}}\rho d\rho \wedge d\phi\\ \\ &
=-{ \frac 12}\sin\theta d\theta \wedge d\phi \hspace*{4.5mm} \end{array}
\end{equation} as in (2.17).

In general, we obtain

\begin{equation} \Omega_n={\frac 12}n \sin\theta d\phi \wedge d\theta
\end{equation} and the corresponding connection 1-forms on $U_+$ and $U_-$ are

\begin{equation} \begin{array}{cl} A_{n+}&={\displaystyle \frac
{n}{2}}(1+\cos\theta)d\phi\\ \\ A_{n-}&={\displaystyle \frac
{n}{2}}(-1+\cos\theta)d\phi \end{array} \end{equation} respectively.

\section{Monopoles over Riemann surfaces of any genus}
\setcounter{equation}{0} \hspace{.9cm} We construct in this section a canonical
monopole connection on any non trivial $U(1)$ bundle over compact Riemann
surfaces of genus $g$. The connection 1-form and curvature 2-form describing
the
monopoles are expressed in terms of the abelian differential of the third kind
with real normalization and integer weights.

The abelian differential $d\tilde{\phi}_{ab}$ of the third kind is a
holomorphic
1-form on the compact Riemann surface except for poles of residue $+1$ and $-1$
at points $a$ and $b$ respectively, with real normalization, that is with pure
imaginary periods when one circles around the canonical cycles.  From
$d\tilde{\phi}_{ab}$ one can construct the abelian integral
$\tilde{\phi}_{ab}$.
Its real part $G(z,\bar{z},a,b,t)$ is a harmonic univalent function over the
Riemann surface with logarithmic behavior around $a$ and $b$

\begin{equation} \begin{array}{cr} \ln{\displaystyle
\frac{1}{|z-a|}}+\:\mbox{regular terms ,}\\ \\ \ln|z-b|+\:\mbox{regular terms
.}
\end{array} \end{equation}

It is a conformal invariant geometrical object and was explicitly constructed
by
Burnside in [6] using a Schottky uniformization of the Riemann surface. $z$
denotes local coordinates on the Riemann surface, and $t$ the set of $3g-3$
parameters describing the moduli space of Riemann surfaces.

Let $a_i\: \:,i=1,\cdots,m$ be $m$ points over the compact Riemann surface, we
associate to them integer weights $\alpha_a \: \:,i=1. \cdots,m$, such that

\begin{equation} \sum_{i=1}^m{\alpha_i}=0 \end{equation}

We define

\begin{equation} \phi=\sum_{i=1}^m{\alpha_i}G(z,\bar{z},a_i,b,t) \: \:.
\end{equation} $\phi$ defines a Morse function over the Riemann surface, it is
exactly the light cone time in the formulation of string theory.

\begin{eqnarray*} \phi \longrightarrow -\infty &\mbox{at the points $ a_i $
with
negative weigths  and }\\ \\ \phi \longrightarrow +\infty &\mbox{at the points
$a_i$ with positive weigths.\hspace*{30pt}} \end{eqnarray*}

$\alpha_i$ are integers in order to have univalent transition functions over
the
non-trivial fiber bundle we will consider.  It corresponds to the Dirac
quantization condition of the magnetic charge.

Let us consider now a $\phi=cte$ curve over the Riemann surface. It is a closed
curve homologous to zero. It divides the Riemann surface into two regions $U_+$
and $U-$, where $U_+$ contains all the points $a_i$ with negative weights and
$U_-$ the ones with positive weights.

We now generalize the connection 1-form (2.35). We define over $U_+$ and $U_-$
the connection 1-forms

\begin{equation} \begin{array}{cr} A_+  ={\frac 12}(\:\:1+\tanh\phi)({\frac
12}\tilde{\phi}_z dz-{\frac 12}\bar{\tilde{\phi}}_{\bar{z}} d\bar{z} ) &
={\frac
12}(\:\:\:1+\tanh\phi)d(Im\tilde{\phi})\\ \\ A_-  ={\frac
12}(-1+\tanh\phi)({\frac12}\tilde{\phi}_z dz-{\frac
12}\bar{\tilde{\phi}}_{\bar{z}} d\bar{z} ) & ={\frac
12}(-1+\tanh\phi)d(Im\tilde{\phi}) \end{array} \end{equation} respectively.

$\tilde{\phi}$ denotes as before the abelian integral whose real part is
$\phi$,
$\tilde{\phi}_z$ denotes its derivative with respect to a local coordinate $z$
over the Riemann surface. $A_+$ and $A_-$ are globally defined in $U_+$ and
$U_-$ respectively and regular therein since the singularity coming from the
factor

\begin{equation} d(Im \tilde{\phi}) \end{equation} at $a_i,\: i= 1,\cdots ,m$,
is cancelled by the corresponding coefficient

\begin{equation} (\pm 1+ \tanh\phi) \end{equation} in the same way as the
factor

\begin{equation} (\pm 1+ \cos\theta) \end{equation} cancelled the singularities
at $\theta=0$ and $\pi$ in (2.35).

In the overlapping $U_+ \cap U_-$ we have

\begin{equation} A_+=A_- + d(Im\tilde{\phi})\:\:. \end{equation}

We now discuss the term in $Im\tilde{\phi}$ more explicitly. There is a one to
one correspondence between $(LC)$ light cone diagrams and the moduli space of
punctured Riemann surfaces, and associated to each $LC$ diagram with given
weights there is a unique abelian differential $d\tilde{\phi}$. It is then
enough to discuss all possible transition functions on the fiber bundle in
terms
of $LC$ diagrams. In the $LC$ diagrams the punctures $a_i$ with positive
weights
are at the right of the diagrams, corresponding to $\phi \longrightarrow
+\infty$, while the $a_i$ with negative weights are at the left of the diagram,
corresponding to $\phi \longrightarrow -\infty$.  For example in the following
figure we have six punctures $a_i$, $i=1,\cdots ,6$, four of them with positive
weights. We denote them $\alpha_1, \alpha_2, \alpha_3 ,\alpha_4 > 0$ .

diagram of the figure corresponds to a torus with 6 punctures, see figure 2.

are integers satisfying $\sum_{a=1}^6{\alpha_i}=0$. The curves $\phi=cte$.
correspond to vertical lines with the corresponding identification of the end
points.

At $\phi= C_1$ the transition function is $e^{in\varphi}$ with

\begin{equation} n=\alpha_1+\alpha_2+\alpha_3+\alpha_4 \end{equation} and
$\varphi$ is the angular variable describing the circle $\phi= C_1$.

If instead we consider $\phi=C_2$ the overlapping between $U_+$ and $U_-$
occurs
at three circles, one enclosing $\textstyle a_3$ and $\textstyle a_4$, another
enclosing $\textstyle a_2$ and the third enclosing $\textstyle a_1$. The
transitions are now $e^{i(\alpha_3+ \alpha_4)\varphi}$ , $e^{i\alpha_2
\varphi}$
and $e^{i\alpha_1 \varphi}$ respectively,where $\varphi$ describes the angular
coordinate at each circle.

We do not consider $\phi=cte$ curves like $\phi=A$ in figure 1,which correspond
to two circles once the identifications of the end points are performed,
because
the transition function on each circle are not necesarily univalent.

The curvature 2-form associated to the connection 1-form (3.4) is given by

\begin{equation} \begin{array}{lll} F= & {{\displaystyle \frac
14}}{\displaystyle\frac{\tilde{\phi}_z \bar{\tilde{\phi}_{\bar{z}}} d\bar{z}
\wedge dz} {\cosh^2\phi}}\:\: = &{{\displaystyle \frac 14}}{\displaystyle
\frac{d\bar{\tilde{\phi}} \wedge d\tilde{\phi}}{\cosh^2\phi}} \end{array}
\end{equation}

Let us now consider the particular case of the Riemann sphere $S_2$, with two
punctures $a$ and $b$ with weights $n$ and $-n$ respectively. Without loss of
generality we can take them to be at $z=\infty$ and $z=0$ of the complex plane
$C\hspace{-7.5pt}l$.   In this case we have

\begin{eqnarray*} \phi= n\ln|z| \end{eqnarray*} We rewrite

\begin{eqnarray*} z=\rho e^{i\varphi} \end{eqnarray*} and use (2.6) to go to the
coordinates on the sphere. We obtain

\begin{eqnarray*} \tanh\phi=\frac{\rho^2-1}{\rho^2+1} = \cos\theta \\ \\
\tilde{\phi}= n\ln{z} \hspace*{2.0cm} \\ \\ Im\tilde{\phi} = n\arg{z} = n\varphi
\:
, \end{eqnarray*} (3.4) then exactly agrees with (2.35) and (3.10) with (2.34).

We have then constructed a canonical monopole connection on non-trivial $U(1)$
fiber bundles on Riemann surface of any genus. The Chern class $c_1$ of the
fiber bundle is determined by the sumation of positive integer weights at the
punctures.

These are all the non trivial bundles that can be constructed over a Riemann
surface.  In fact, any complex vector bundle over an open Riemann surface is
trivial , hence any complex vector bundle over $U_+$ or $U_-$ is trivial and we
have constructed all the posible transitions between fibers over $U_+ \cap
U_-$.

\section{ Seiberg-Witten equations over Riemann surfaces}
\setcounter{equation}{0} \hspace{.9cm}We obtain in this section the reduction
of
the Seiberg - Witten 4-monopole equations [7] to  2-dim Riemann surfaces . We
then show that the canonical connections over compact Riemann surfaces we
constructed in section 3 are solutions to these equations. It is convenient to
use complex notation

\begin{equation}
x^{\mu}\sigma_{\mu}=z\sigma_{z}+\bar{z}\sigma_{\bar{z}}+y\sigma_{y}+\bar{y}\sigma_{\bar{y}}
\end{equation} Where \begin{equation} \begin{array}{cc}
¥\sigma_{z}=\left(\matrix{1 &0 \cr \cr 0 &0}\right) &
\sigma_{\bar{z}}=\left(\matrix{0 &0 \cr \cr 0 &1}\right) \\ \\
¥\sigma_{y}=\left(\matrix{0 &0 \cr \cr i &0}\right) &
\sigma_{\bar{y}}=\left(\matrix{0 &i \cr \cr 0 &0}\right) \end{array}¥
\end{equation} The 4-monopole equations [7] are
\renewcommand{\theequation}{\arabic{section}.\arabic{equation}a}
\begin{eqnarray}
 F^{\mu\nu}\sigma_{\mu\nu\alpha\beta}=&\bar{M}_{(\alpha}
M_{\beta)} \end{eqnarray}
\renewcommand{\theequation}{\arabic{section}.\arabic{equation}b}
\setcounter{equation}{2} 
\begin{eqnarray} \sigma^{\mu}{\cal D}_{\mu}M=&0
\end{eqnarray} 
\renewcommand{\theequation}{\arabic{section}.\arabic{equation}}

We consider the reduction to 2-dim in an open neighborhood by taking
$\partial_y
= \partial_{\bar{y}} = 0$ and then extend the resulting equations to compact
Riemann surfaces. We obtain after the reduction of (4.3a)

\begin{eqnarray}
 -F^{z\bar{z}}&=&M^{\ast}_1M_1-M^{\ast}_2M_2 \\ \nonumber \\
 F^{y\bar{z}}&=&M_2M^{\ast}_1 
 \end{eqnarray}
 ¥ Where $F^{y\bar{z}}
= F_{\bar{y}z} = -\partial_{z}A_{\bar{y}}$, and $\ast$ denotes complex
conjugation, for (4.3b) we get
\renewcommand{\theequation}{\arabic{section}.\arabic{equation}a}
\begin{equation} {\cal D}^{z}M_1+iA_{\bar{y}}M_2=0` \end{equation}
\renewcommand{\theequation}{\arabic{section}.\arabic{equation}b}
\setcounter{equation}{5} \begin{equation} ¥{\cal D}^{\bar{z}}M_2+iA_{y}M_1=0
\end{equation} \renewcommand{\theequation}{\arabic{section}.\arabic{equation}}
We now look for a conformal extension over a compact Riemann surfaces. From
(4.4) $M_1$ and $M_2$  must transform as one forms under a change of chart.
From
(4.5) $A_{\bar{y}}$ must consequently change also as a one form. However this
change is not compatible with (4.6) unless \begin{equation}
\partial_{z}A_{\bar{y}}=0 \end{equation}¥ Which implies

\begin{equation} M_2 M^{\ast}_1=0 \end{equation}

We end up with the following set of equations \begin{equation}
-F_{\bar{z}z}=M^{\ast}_1M_1-M^{\ast}_2M_2 \end{equation}¥ and
\renewcommand{\theequation}{\arabic{section}.\arabic{equation}a}

\begin{equation} {\cal D}_{\bar{z}} M_1=0 \: \: M_2=0 \\ \\ \end{equation} or
\renewcommand{\theequation}{\arabic{section}.\arabic{equation}b}
\setcounter{equation}{9}

\begin{equation} ¥ {\cal D}_{z} M_2=0 \: \: M_1=0\: , \end{equation}
\renewcommand{\theequation}{\arabic{section}.\arabic{equation}}

Where $M_2d\bar{z}$ $(M_1dz)$ is a one form over the compact Riemann surfaces.
We will call equations (4.9) and (4.10) the monopole equations over Riemann
surfaces. We will compare (4.9), (4.10) with the Hitchin equations over compact
Riemann surfaces, which arise from a reduction of the $SU (2)$ selfdual
equations in 4 dim, in a next communication.

We now show that the canonical connections we constructed in section 3 give
rise
to a class of regular solutions to (4.9), (4.10). We consider the non trivial
$U(1)$ bundle over a compact Riemann surfaces of genus g, with transitions as
indicated in the figure 3. 
 If we take the transitions as indicated in the diagram
the transition functions are single valued over the compact Riemann surfaces as
a consequence of our assumption of integer weights $\alpha_i$, $i= 1,\cdot,m$,
$\sum_{i=1}^{m}\alpha_i=0$, at the points $\textstyle a_i$. The canonical
connection of section 3 can now be reexpressed as:

\begin{equation} \begin{array}{cll} A_{+z}=&{\displaystyle {\frac
12}}(1+\tanh({\phi})) \tilde{\phi}_z dz &\mbox{in $U_+$ ,}\\ \\
A_{z}=&{\displaystyle {\frac 12}}\tanh{(\phi)} \tilde{\phi}_z dz \hspace*{40pt}&
\mbox{in $U$ ,} \\ \\ A_{-z}=&{\displaystyle {\frac 12}}(-1+\tanh({\phi}))
\tilde{\phi}_z dz & \mbox{in $U_-$ .} \end{array}¥ \end{equation}

It is easiest to solve (4.9) and (4.10) first at U.   We obtain

\begin{eqnarray} M_2=h(\phi)\partial_{\bar{z}}\bar{\tilde{\phi}}, ,\nonumber \\
\\ {\displaystyle {\frac 12}}\partial_z \tilde{\phi}\:h^{\prime} + A_z h=0 \:
,\nonumber \end{eqnarray}

We thus have for $h\neq 0$,

\begin{eqnarray} A_z&=&-{\displaystyle {\frac
12}}\partial_{z}{\tilde{\phi}}\:{\displaystyle {\frac {h^{\prime}}{h}}}
\nonumber \\ \\ F_{\bar{z}z}&=&-{\displaystyle {\frac
14}}\partial_{\bar{z}}\bar{\tilde{\phi}}\:\partial_{z}{\tilde{\phi}} \ln( h
\bar{h} )^{\prime \prime} \nonumber \end{eqnarray}¥
where $\partial_{z}{\tilde{\phi}}={\tilde{\phi}}_z$.

We finally obtain
\begin{eqnarray} h={\displaystyle {\frac {1}{\sqrt{2} \cosh{\phi}}}}&\mbox{ in
$U$ ;} \end{eqnarray}

We now go back to (4.11) and notice that the solutions are

\begin{eqnarray} h={\displaystyle {\frac {\exp{-\hat{\phi}}}{\sqrt{2}
\cosh{\phi}}}}&\mbox{ in $U_+$ ;} \end{eqnarray}

and

\begin{eqnarray} h={\displaystyle {\frac {\exp{\hat{\phi}}}{\sqrt{2}
\cosh{\phi}}}}&\mbox{ in $U_-$ ;} \end{eqnarray}¥ where $\hat{\phi}= Im
\tilde{\phi} $

We notice that h is a single valued object at $U_+$ and $U_-$ in spite of the
presence of $\hat{\phi}$ which is multi-valued.This is so because of the
integer
weights we have considered. $\exp{\hat{\phi}}$ is not single valued in U,when
we
go around a handle. This is the reason why we have considered the transition as
in figure 3.

We could construct a similar solution in terms of $M_2$. The only change is the
reverse in the sign of $A_z$. These are monopoles with opposite magnetic
charge.

We have thus constructed a class of regular solutions to the monopole equations
over Riemann surfaces of genus g.   The class of solutions depend on the
conformal class of the given Riemann surfaces as well as in m points over the
surface with integer weights $\alpha_i$,$i= 1,\cdot,m$,
$\sum_{i=1}^{m}\alpha_i=0$. The Chern class of the fiber bundle is given by
$(\pm 1)$ times the sum of the positive weights.

\section{ Conclusions} \hspace{.9 cm}We obtained an explicit canonical
construction of monopoles solutions for any non trivial bundle over any compact
Riemann surface of genus g. The solutions depend on the conformal class of the
Riemann surface and on a set of integer weights. We gave the explicit solutions
for the Weyl spinors of the Seiberg-Witten monopole equations on the non trivial bundle
and reduced them onto compact Riemann surfaces of any genus g. It is
interesting
to pointout
that the same reduction but for the $SU(2)$ self-dual equations over 4
dimensions yields the 
Hitchin's equations [8] over Riemann surfaces.The latter have a similar
structure to
the equations 
(4.9),(4.10) but in terms of Higgs fields which behave as  1-forms over the
Riemann surface 
instead of the Weyl spinors. This distinction makes an important differenc
between the two sets of equations. In particular, the Dirac monopole is \underline {not} 
a solution of Hitchin's equations [9] but it is a solution of the reduced Seiberg-Witten equations.


\begin{thebibliography}{\large References} \bibitem{}N.Seiberg and E.Witten,
{\it Nucl. Phys. }{\bf B426} (1994) 19; Erratum, {\bf B430 } (1994) 485; {\bf
B431 } (1994) 484. \bibitem{}P.Goddard and D.Olive {\it Rep. Prog. Phys. }{\bf
41 } (1978) 1357. 
\bibitem{}C.H.Taubes, Preprint Harvard University:"The
Seiberg-Witten and the Gromov invariants"(1995); {\it Math. Res. Lett.}{\bf1}
(1994) 809. 
\bibitem{}I.Martin and A.Restuccia {\it Phys.Lett.}{\bf B271}(1991) 361;\\ M.LLedo, I.Martin, A.Restuccia and A.Mendoza {\it Lett. Math.
Phys.}{\bf 24} (1992) 275. 
\bibitem{}A.Trautman {\it Int. Jour. Theor.Phys.}{\bf 16} (1977) 561. 
\bibitem{}W.Burnside {\it Proc. London Math.
Soc.}{\bf 23} (1881) 49.  
\bibitem{}E.Witten {\it Math.Res. Lett.}{\bf 1} (1994) 769. 
\bibitem{}N.Hitchin{\it Proc. London Math.
Soc.}{\bf 55} (1987) 59;Lectures al LASSF 89, Caracas, Venezuela.
\bibitem{}I.Martin, R.Martinez and A. Restuccia, preprint USB/SF/96-236.
\end{thebibliography}
\end{document}